\documentclass[aps,prc,a4paper,twocolumn]{revtex4}
\usepackage{graphicx}
\begin{document}

\title{Nodal Lines in the  Cranked HFB Overlap Kernels}
\author{Makito Oi}
\email[e-mail: ]{ m.oi@surrey.ac.uk}
\affiliation{Department of Physics, University of Surrey,\\ 
Guildford, GU2 7XH, United Kingdom}
\author{Naoki Tajima}
\email[e-mail: ]{ tajima@apphy.fukui-u.ac.jp}
\affiliation{Department of Applied Physics, Fukui
University, \\3-9-1 Bunkyo, Fukui 910-8507, Japan}

\date{26-Oct-2004}

\begin{abstract}
Norm overlap kernels of the cranked Hartree-Fock-Bogoliubov states
are studied in the context of angular momentum projection.
In particular, the geometrical distribution of nodal lines,
i.e., one dimensional structures where the overlap kernels possess
null value, is investigated 
in the three dimensional space defined by the Euler 
angles. It is important to know 
the distribution of these nodal lines when one attempts to
determine the phase of norm overlap kernels.
\end{abstract}
\maketitle
\addvspace{7mm}

The mean field approximation (MFA)  is  successful in the description of 
nuclear systems \cite{BHR03,Fr01,RS80}, to which the independent
particle motion picture can be applied \cite{RS80,Br54,BG57}. 
Deformation  and superfluidity (superconductivity) 
in nuclei are good examples \cite{RS80}
that are explained by MFA.
Although some essential correlations are taken into
account through MFA in an efficient manner, it fails to incorporate
 other important features such as symmetries 
and relevant higher order correlations,
which may be necessary for studying nuclear structures further
\cite{BHR03,RS80}.

Within MFA,  
physical quantities such as energy and quadrupole moment are 
calculated as expectation values, but there is room
for further quantisations. For example,
angular momentum is not conserved when the rotational symmetry
is spontaneously broken by  deformed mean fields.
As a consequence, the corresponding
many-body state has the form of a wave packet \cite{IMR79}.
In the cranking model \cite{Ing56}, which describes a rotating mean field 
in a semi-classical manner, the mean field state is written as
\begin{equation}
|\psi_{\rm MFA}(\omega)\rangle = \sum_{IM}C^I_{M}(\omega)|IM\rangle.
\end{equation}
(The rotational frequency of the deformed mean field is denoted by
$\omega$, while the quantum numbers for 
total angular momentum and its magnetic quantum number
are expressed as $I$ and $M$, respectively
\footnote{In this paper, we denote angular momentum operators
by $\hat{J}_i$ ($i=1,2,3$); corresponding quantum numbers
by $I$ and $M$; and the expectation values by
$J_i=\langle\hat{J}_i\rangle$.
$I$ and $M$ are integers while $J_i$ are real numbers and not
necessarily integers.}.)
$|\psi_{\rm MFA}(\omega)\rangle$ usually has large  fluctuations 
in its probability distribution $|C^I_M|^2$. 
The width of the fluctuations becomes larger at higher spin \cite{IMR79},
which means $|\Psi_{\rm MFA}(\omega)\rangle$ 
possesses a more averaged character.
As a consequence, a simple application of the cranking model
faces  difficulty, for instance, in the description of
the mixture of two different states 
in band crossing regions \cite{Ham76}.

A quantum mechanical description of collective rotation is given
in the generator coordinate method (GCM) \cite{HW52} 
with a choice of the Euler angles
($\Omega \equiv (\alpha,\beta,\gamma$)) as generator coordinates.
This method corresponds to angular momentum
projection (AMP) \cite{PY57}. Rotational symmetry
spontaneously broken by the deformed mean field can be restored 
by superposing the states that point various orientations
$|\psi(\Omega)\rangle \equiv \hat{R}(\Omega)|\psi _{\rm MFA}\rangle$ .
( $\hat{R}(\Omega )$ is a
rotational operator in the three dimensional space.
See Eq.(\ref{rot_op}) below.)
A state with  symmetry restoration $|\Phi\rangle$ is 
schematically expressed as
\begin{equation}
|\Phi\rangle = \int d\Omega f(\Omega)|\psi(\Omega)\rangle,
\end{equation}
with $f(\Omega)$ being the weight function that is determined
by the variational  principle.
This state $|\Phi\rangle$, a superposition of the infinite number of 
degenerate states,  corresponds to a Nambu-Goldstone 
mode \cite{NG60} in a finite system.

\begin{figure*}[tpbh]
\begin{tabular}{cc}
\includegraphics[height=0.21\textheight,width=0.5\textwidth]{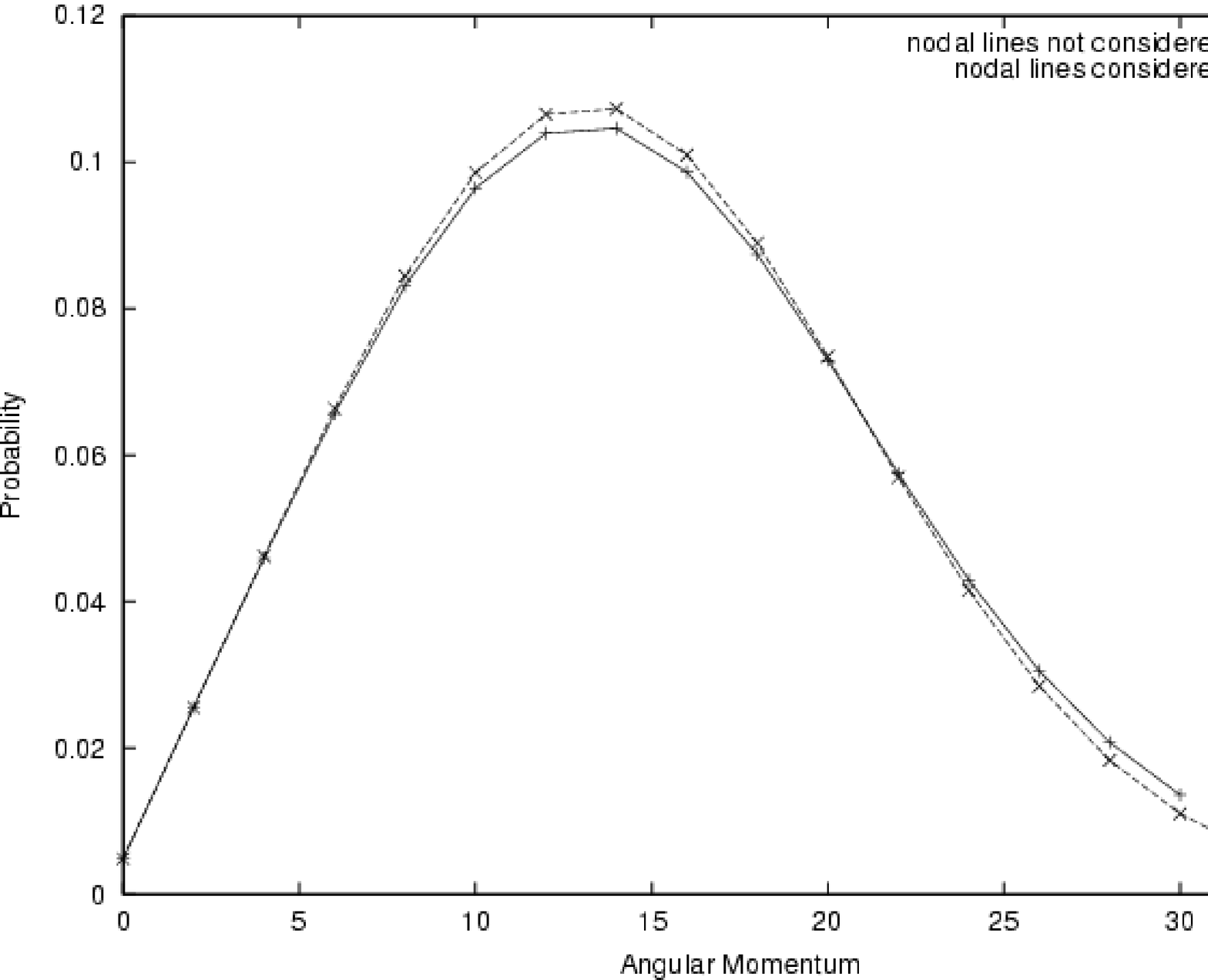}&
\includegraphics[height=0.21\textheight,width=0.5\textwidth]{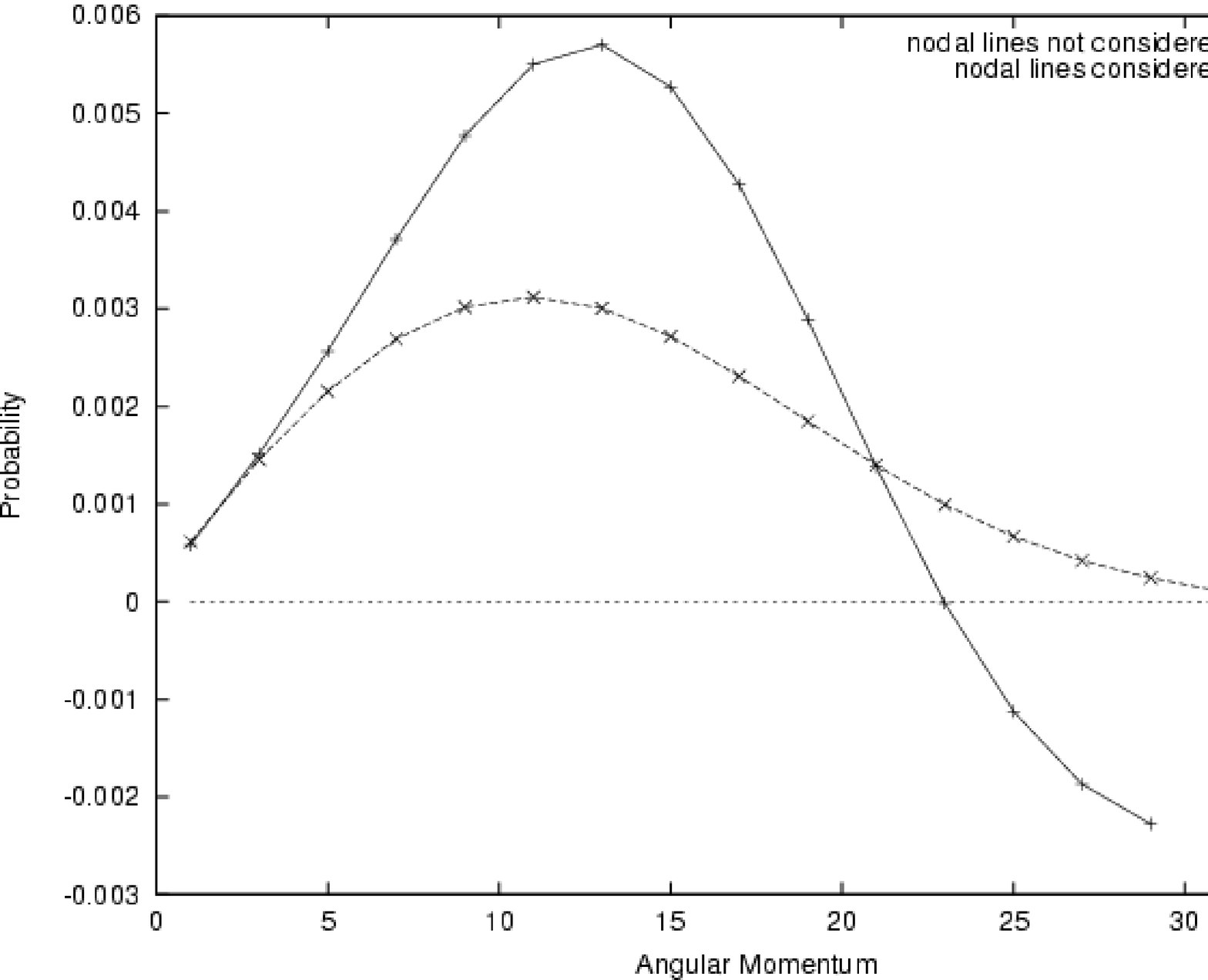}\\
\hline
Even-$I$ & Odd-$I$.\\
\hline
\end{tabular}
\caption{Probability distributions of angular momentum using
two types of prescription to determine the sign of the norm overlap
 kernels. The solid line
represents the probability obtained by the old method while the dashed
line is by the new method.  
 The constraint value on angular momentum is $J=14\hbar$.
The left  (right) panel shows even-$I$ (odd-$I$) components.
Note the different  scales in the first and second panels.}
\label{fig_prob}
\end{figure*}

In our previous works,
we have numerically performed  AMP 
in  attempts to describe high-spin states, such as
tilted rotation for high-$K$ states in $^{178}$W  \cite{OWA02} 
and the wobbling motion in the multi-band crossing region 
in $^{182}$Os 
\cite{OAHO00}. 
Through these works, we found
it difficult to perform numerical calculations of AMP in some situations.
Such situations are observed to occur, for example, 
in the calculations of AMP applied to cranked HFB states 
at high spin (angular momentum ranging from $10$ to $20\hbar$).
A typical problem is negative values of the probability
$|C^I_M|^2$ for $I$ being odd integer,
although the probabilities for even $I$ seem reasonably calculated
\footnote{Odd-$I$ components are much smaller (less than $1\%$ for each
odd-$I$) than even-$I$ components but do not vanish completely.
Despite the conservation of the signature quantum number,
the simultaneous presence of even and odd-$I$ components is 
observed in the calculations.
It is due to the gamma-deformation caused by 
the self-consistent cranking calculations, as explained in \cite{OOTH98},
which mix states having non-zero $K$ quantum number.}.
(See Fig.\ref{fig_prob}.) 

We first suspected the coarse mesh  employed
in the numerical integration causes this difficulty,
because the integration in the AMP formula (See Eq.(\ref{eq_ovrlp}) below) 
is discretised in numerical calculations with respect to the Euler angles.
The appropriate mesh size can be estimated by using
the uncertainty principle
$\Delta I \Delta \Omega \simeq \hbar$.
The constraint value $\langle\hat{J}_1\rangle = J$ and the fluctuation
$\Delta I$ are roughly related as $\Delta I \simeq J/2$ at high spin 
\cite{IMR79,OOTH98}.
Then, $\Delta\Omega =2\hbar/J$. For example, $\Delta\Omega \simeq
6^{\circ}$ for $J=20\hbar$. 
The appropriate number of mesh points 
are therefore about 60 for $\alpha, \gamma,$ 
and 30 for $\beta$.
Even with this mesh size, however, the problem was observed to occur 
in performing AMP 
when high spin states 
 are considered (i.e.,$J > 10\hbar$).
On the other hand, when the angular momentum is low ($J\le 10\hbar$), AMP
is successfully performed, as expected.

We then suspect that the poor determination of the sign of
the norm overlap kernels, which will be given in Eq.(\ref{ons_frml}),
may cause this problem.
These kernels are necessary in the AMP procedure, in which the
state in the projected space is expressed as
\begin{equation}
\label{proj_state}
|\Psi^I_M\rangle=\sum_K g^I_K\hat{P}^I_{MK}|\psi_{\rm MFA}\rangle.
\end{equation}
The coefficients $g^I_K$ are determined by
a variational equation
\begin{equation}
\sum_{K}\left(H^I_{MK}-E^IN^I_{MK}\right)g^I_K=0.
\end{equation} 
Here, $\hat{P}^I _{MK}$ denotes the AMP operator, and matrices 
$N^I_{MK}$ and $H^I_{MK}$ are defined as,
\begin{eqnarray}
\label{eq_ovrlp}&&\left(
\begin{array}{c}
N^I_{MK} \\ H^I_{MK}
\end{array}
\right)
=\langle\psi_{\rm MFA}|
\left(
\begin{array}{c}
 \hat{P}^I_{MK}\\ \hat{H}\hat{P}^I_{MK}
\end{array}
\right)
|\psi_{\rm MFA}\rangle\\ \nonumber
&=&\!\!\frac{2I+1}{8\pi^2}\!\! \int \!\! d\Omega D^{I*}_{MK}(\Omega)
\langle\psi_{\rm MFA}|
\left(
\begin{array}{c}
 \hat{R}(\Omega)\\ \hat{H}\hat{R}(\Omega)
\end{array}
\right)|\psi_{\rm MFA}\rangle.
\end{eqnarray}
%
A useful relation to note here is $|C^I_M|^2 = N^I_{MM}$.
The measure in the three-dimensional Euler space $d\Omega$ 
is defined as $d\alpha \sin\beta d\beta
d\gamma$, and the integration intervals
 are $\left[0,2\pi\right]$ for $\alpha$ and
$\gamma$, and $\left[0,\pi\right]$ for $\beta$.
The rotational operator $\hat{R}(\Omega)$ is given as
\begin{equation}
 \label{rot_op}
\hat{R}(\Omega)=\exp(-i\alpha\hat{J}_z)\exp(-i\beta\hat{J}_y)\exp(-i\gamma\hat{J}_z).
\end{equation}

The quantities $\langle\psi_{\rm MFA}| \hat{R}(\Omega)|\psi_{\rm MFA}\rangle$ 
and $\langle\psi_{\rm MFA}| \hat{H}\hat{R}(\Omega)|\psi_{\rm
MFA}\rangle$ in Eq.(\ref{eq_ovrlp})
are called norm and energy overlap kernels, respectively.
They are calculated through formulae derived by
Onishi et al \cite{OY65,OH80}.
By utilising the Onishi formulae, one can calculate the norm overlap kernel as
\begin{equation}
N(\Omega)=\langle\psi_{\rm MFA}|\hat{R}(\Omega)|\psi_{\rm MFA}\rangle = 
\sigma(\Omega)\sqrt{{\rm det}P(\Omega)},
\label{ons_frml}
\end{equation}
where $\sigma(\Omega)$ has values $\pm 1$.
This ambiguity in sign comes from the square root operation.
We will come back to this point shortly.
The matrix $P(\Omega)$ is written \cite{OH80},
\begin{equation}
 P(\Omega)=U^{\dag}D^{\dag}(\Omega)U+V^{\dag}D^{T}(\Omega)V,
\end{equation}
where
$U$ and $V$ are matrices defined in the
general Bogoliubov transformation \cite{RS80} between canonical
($c_{\mu} \ c^{\dag}_{\mu}$) and quasi-particle
annihilation and creation operators 
$(\beta_{i} \ \beta^{\dag}_{i}$):
\begin{eqnarray}
\left(
 \begin{array}{c}
 \beta_i\\
 \beta_i^{\dag}
  \end{array}
\right)
&=&
\left(
\begin{array}{cc}
U_{\mu i}^* & V_{\mu i}^* \\
V_{\mu i} & U_{\mu i} 
\end{array}
\right)
\left(
 \begin{array}{c}
 c_{\mu}\\
 c_{\mu}^{\dag}
  \end{array}
\right).
\end{eqnarray}

The presence of $\sigma(\Omega)$ implies that
${ N}(\Omega)$ is a two-valued function of $\Omega$
although it should be physically well-defined and single-valued.
Therefore, it is necessary to choose an appropriate sign
for given $\Omega$. 
Two methods have been proposed \cite{HHR82,NW83} to determine the
sign of $\sigma(\Omega)$, but due to its numerical feasibility the method
of Hara, Hayashi and Ring \cite{HHR82} is widely used. It makes use of
the continuity of the norm overlap kernel as a function of the Euler
angles $\Omega$. Our calculations are
based on this method, and the procedure is explained below.

First of all, the origin in the Euler space is
assumed to have some certain phase. (In our case, $\sigma(0)=+1$.)
Then, by analytical continuation,
the well-defined region is extended 
until the whole Euler space is covered.
To achieve this, the Euler space is divided into domains
defined as follows:
Writing $N(\Omega)=r(\Omega){\rm e}^{i\theta(\Omega)}$, 
the argument $\theta(\Omega)$ and 
the norm $r(\Omega)$ are supposed to be continuous
with respect to $\Omega$.
Then,  a point $\Omega=\Omega(\alpha,\beta,\gamma)$ 
belongs to a domain $D_n \ (n=0,\pm 1,\pm 2, \cdots)$ when
$(n-\frac{1}{2})\pi < \theta(\Omega)\le (n+\frac{1}{2})\pi$.
Inside each domain, 
$\sigma(\Omega)$ is determined  to be $+1$ for even $|n|$ 
and $-1$ for odd $|n|$.
The boundary between $D_n$ and $D_{n+1}$ is determined by
tracking the continuity of $\theta(\Omega)$ through the
derivative information.
In this way, ${N}(\Omega)$ becomes well-defined as
 a single-valued function in the entire domain $D\equiv \bigcup_{n}D_n$
(i.e., the whole Euler space) .

As mentioned above, derivatives of the norm overlap kernels are 
used to check the continuity of $\theta(\Omega)$.
A set of formula derived by Onishi et al. is again useful
to calculate these derivatives \cite{OH80},
which are now given in the form of logarithmic derivatives,
\begin{equation}
\label{OnsDrv}
\frac{\partial}{\partial \Omega}\ln \langle\psi_{\rm
MFA}|\hat{R}(\Omega)|\psi_{\rm MFA}\rangle =\frac{1}{2}\frac{1}{{\rm
det}P(\Omega)}\frac{\partial}{\partial \Omega}{\det}P(\Omega).
\end{equation}
Detailed forms for the formulae are given in Ref.\cite{OH80}.
(Sometimes the formulae are called generalised Wick's theorem \cite{RS80}.)
Because the argument of the norm overlap kernel is written as,
\begin{equation}
\theta(\Omega) = {\rm Im}\left[ 
\ln\langle\psi_{\rm MFA}|\hat{R}(\Omega)|\psi_{\rm MFA}\rangle\right],
\end{equation}
the derivatives of $\theta(\Omega)$ can be computed through Eq.(\ref{OnsDrv}).
Note that the derivative of the argument, 
$\theta'= \frac{\partial \theta}{\partial \Omega}$,
is uniquely evaluated , i.e., being free from the sign ambiguity,
because these formulae essentially 
make use of the square of the norm overlap kernels
(${ N}(\Omega)^2$) to eliminate the ambiguity coming from $\sigma(\Omega)$.

To specify the domain $D_n$,
we consider the following two quantities
that can be calculated from a pair of neighbouring mesh points displaced by
small distance $\Delta\Omega$, that is,
(i) the difference between the two points, corresponding to
the approximate derivative of the argument $\theta(\Omega)$ obtained
by the two-point formula, which is 
\begin{equation}
  a(\Omega)\equiv\frac{\theta(\Omega+\Delta\Omega)-\theta(\Omega)}
  {\Delta\Omega}
\end{equation}
and (ii) the average of $\theta'$ obtained through Eq.(\ref{OnsDrv}) 
between the two points, which is
\begin{equation}
  b(\Omega)\equiv\frac{\theta'(\Omega+\Delta\Omega)+\theta'(\Omega)}{2}.
\end{equation}
Note that only $a(\Omega)$ suffers from the sign ambiguity,
but $b(\Omega)$ does not.
We can now determine 
the relative sign of the kernel between the points $\Omega$ and
$\Omega + \Delta \Omega$ 
by checking the magnitude of the residue 
$\delta\theta(\Omega)\equiv a(\Omega)-b(\Omega)$.
When the value of the residue is small, or behaves as
$\delta\theta(\Omega) \simeq -\frac{\theta^{(3)}(\bar{\Omega})}{12}\Delta\Omega^2$ 
with $\Omega < \bar{\Omega} < \Omega+\Delta\Omega$,
the neighbouring point $\Omega+\Delta\Omega$ is judged to be in the same 
domain $D_n$ as the reference point $\Omega$. 
On the contrary, 
the point $\Omega+\Delta\Omega$ belongs to the different domain 
if the following two conditions are satisfied:
(i) $\delta\theta$ returns a significant deviation from the 
behaviour of $-\frac{\theta^{(3)}(\bar{\Omega})}{12}\Delta\Omega^2$;
and (ii) the deviation is removed after the sign $\sigma(\Omega+\Delta\Omega)$
is inverted and the resultant residue follows the behaviour 
of $-\frac{\theta^{(3)}(\bar{\Omega})}{12}\Delta\Omega^2$.

According to the results in our calculations,
the above naive approach works for $J\le 10\hbar$, 
but it seems to fail when  high spin states 
are considered ($J > 10\hbar$).
A typical symptom of the failure 
is seen as violation of the positive definite nature in probability,
as illustrated in the right portion of Fig.1.
In Fig.1,
angular momentum components in 
the cranked HFB states at $J=14\hbar$ are plotted, that is,
\begin{equation}
  W^I\equiv \text{Tr}(N^I) 
  = \sum_{M=-I}^{I}|C^I_M(\langle\hat{J}_1\rangle=14\hbar)|^2,
\end{equation}
where the matrix $N^I$ is defined in Eq.(\ref{eq_ovrlp}) and 
the trace is taken with respect to the magnetic quantum number $M$.
With the above method to determine the sign (solid lines in the figure),
AMP seems to work properly when the even-$I$ distribution
of $W^I$ is concerned (i.e., the left portion in the figure),
while in the right portion the positive definite nature of $W^I$
is violated slightly to an order of magnitude of $10^{-3}$.
One may say that such small numerical errors can be neglected
when proper physics is extracted from the results.
In fact, as far as the ground state rotational bands are concerned (which 
consist of only even-$I$ components),  it may be justified to 
ignore these small errors.
However, we would like to investigate the cause of the errors 
in this paper because high-spin physics involving odd-$I$
states (e.g., high-$K$ bands) becomes more and more important today.

The above problem (i.e., negative probability) implies that 
the resultant norm overlap kernels do not satisfy proper features
as a physical quantity. A possible reason for this defect is
the assumption of the continuity of $\theta(\Omega)$.
It is, in fact,  singular on a nodal line, 
i.e., a set of points where the norm overlap kernel 
becomes zero ($r(\Omega)=0$).
The analytical continuation approach assumes the
continuity of norm overlap kernels 
(more precisely, of the argument $\theta(\Omega)$),
so that the method fails to work in the neighbourhood of nodal lines.

To deal with the nodal lines in the context of the sign determination, 
there are basically two approaches. One is to find out
all the nodal lines in the Euler space 
(from the information of $r(\Omega)=0$). 
Once the locations of the nodal lines are known,
one can determine $\sigma(\Omega)$ without any ambiguity
because we can avoid the calculations of
the residue $\delta\theta$ between two points sandwiching the nodal lines.
The other is to calculate the residues for all the possible directions
(that is, $\Delta\Omega=$$\pm \Delta\alpha$, $\pm \Delta\beta$, 
and $\pm \Delta\gamma$) no matter where the nodal lines are.
In this case, optimisation regarding to the values of the residues is made 
by fixing $\sigma(\Omega+\Delta\Omega)$ in a manner of trial-and-error.
In this paper, the second approach is chosen for the sake 
of simplicity in numerical calculations.

\begin{figure}[b]
\begin{tabular}{c}
\includegraphics[width=0.45\textwidth]{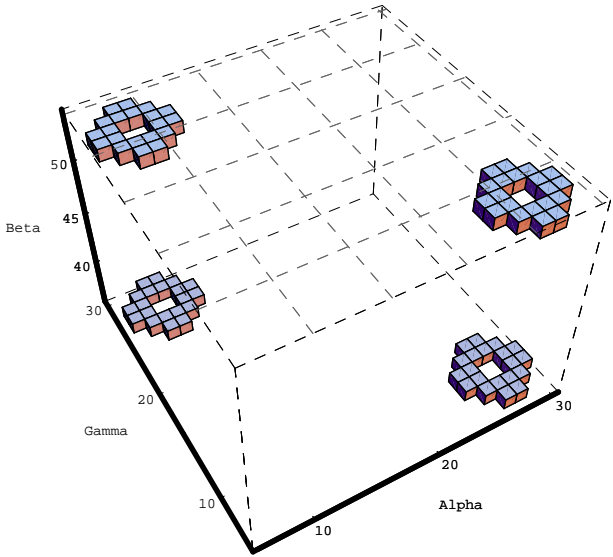}\\
(A) $\Delta \alpha = \Delta\gamma = 10^{\circ}$, and $\Delta\beta=2^{\circ}$.\\
\\
\includegraphics[width=0.45\textwidth]{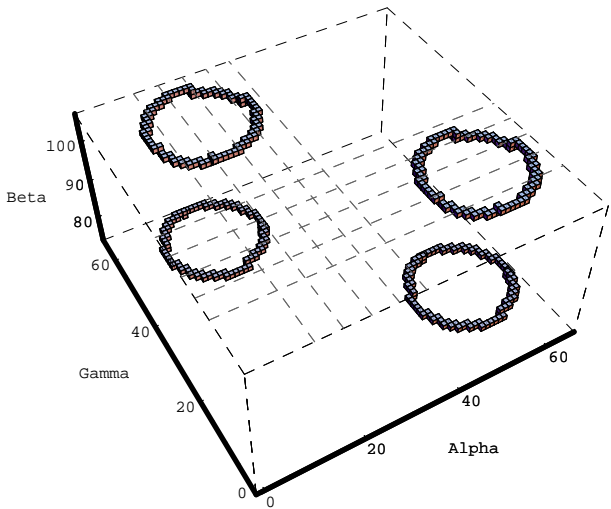}\\
(B)$\Delta \alpha = \Delta\gamma = 5^{\circ}$, and $\Delta\beta=1^{\circ}$.\\
\\
\end{tabular}
\caption{Nodal lines at $J=12\hbar$ with low and high resolutions
for the mesh size.}
\label{fig_nodal1}
\end{figure}

\begin{figure}[b]
\begin{tabular}{c}
\includegraphics[width=0.45\textwidth]{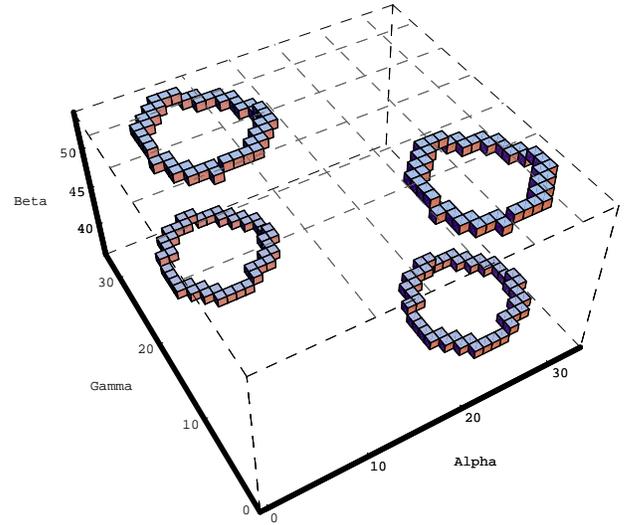}\\
(A) $\Delta \alpha = \Delta\gamma = 10^{\circ}$, and $\Delta\beta=2^{\circ}$.\\
\\
\includegraphics[width=0.45\textwidth]{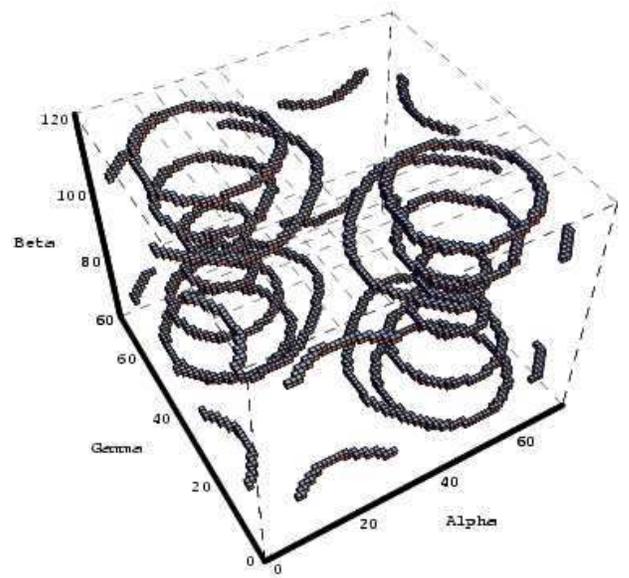}\\
(B) $\Delta \alpha = \Delta\gamma = 5^{\circ}$, and $\Delta\beta=1^{\circ}$.\\
\\
\end{tabular}
\caption{Nodal lines at $J=14\hbar$ with low and high resolutions
for the mesh size. The detailed structure of (B)
is shown in Fig.\ref{fig_nodal4}.}
\label{fig_nodal2}
\end{figure}

Although there is no need to know 
explicit information about the geometry and topology
of the nodal lines in the present approach, we realise that 
such information are very useful to clarify the nature and source
of the problem. Therefore, in this paper,
we try to pin down the locations of nodal lines, and
succeed to find them for the first time.
(Figs.\ref{fig_nodal1} and \ref{fig_nodal2}).

Let us first discuss the nodal lines we found,
before the detailed explanation is given on the new method 
to determine the sign $\sigma$.
To obtain norm overlap kernels, we first perform principal axis cranked HFB
calculations with a constraint on angular momentum $\langle
\hat{J}_1\rangle \equiv J$ (as well as 
constraints on particle numbers). The pairing-plus-Q$\cdot$Q force
is employed with the standard choices for the force parameters and model space
\cite{BK65}. This HFB calculation is performed in a fully self-consistent
manner, so that the gamma (triaxial) deformation is self-consistently handled.
The details of the method used for this calculation are
explained in Ref.\cite{HO96}.
Then, as already mentioned above, 
the overlap kernels are obtained by means of the Onishi formulae
such as Eqs. (\ref{ons_frml}) and (\ref{OnsDrv}).
The modified method for the phase determination, which 
will be explained later, is employed to calculate the norm 
overlap kernels.
$^{170}$Dy is chosen in this study,
as a well-deformed rare-earth nucleus. 
($\beta=0.295, \Delta_p = 0.867$ (MeV), $\Delta_n=0.652$ (MeV)
 at $J=0$ \cite{MN}.) 
The yrast line of this nucleus does not show
any crossings at high spin ($\le 20\hbar$) in our calculation.
(The more complicated structure of nodal lines may appear in the
cranked HFB states in band-crossing regions. Such states are planned to
be studied in future.) 

Figs.\ref{fig_nodal1} and \ref{fig_nodal2} present the nodal lines
found in  cranked HFB states at $J=12\hbar$ and $14\hbar$, respectively.
In the figures, cubes are used to show and cover the regions where nodal lines
exist, for  graphical convenience. The size of the cubes is the same as
the mesh size used in AMP. It is given by 
$\Delta\Omega=L_{\Omega}/N_{\Omega}$, where
$L_{\alpha}=L_{\gamma}=2\pi$ and $L_{\beta}=\pi$.
The number of the mesh points are  $N_{\alpha}=N_{\gamma}=36 \ (72)$ and
$N_{\beta}=90 \ (180)$ for the low (high) resolution calculations.

The positions of nodal lines are determined in the following  approximation:
on each face of the cube,
after the phase determination, the norm kernels are linearly
interpolated with respect to the Euler angles, that is,
\begin{equation}
  \label{linApp}
{ N}_{\text{linear}}
(\Omega)\simeq c_1\alpha + c_2\beta + c_3\gamma + c_4.
\end{equation}
The coefficients $c_i$ are determined by the exact values of ${ N}(\Omega)$
at the four vertices on the face.
The position of the nodal line on the face is
obtained by solving the simultaneous equations
of 
\begin{equation}
\text{Re}[{ N}_{\text{linear}}]=
\text{Im}[{ N}_{\text{linear}}]=0.
\end{equation}

In the present study, no nodal lines are found in $0\le J \le 10\hbar$.
Nodal lines are found in $12\hbar \le J \le 20\hbar$
and form a simple topology of closed loops (rings) without knots. 
Their geometrical distributions  seem to become more complicated  
as the value of $J$ becomes larger.
It is seen in Fig.\ref{fig_nodal2} that some of the nodal lines are missing 
in the low resolution calculation
but this result is due to the simple linear approximation,
Eq.(\ref{linApp}), for the complicated geometrical distribution
of the nodal lines.

Symmetries seen in Figs.\ref{fig_nodal1} and \ref{fig_nodal2},
are expressed by
\begin{eqnarray}
\label{sym1}
&& \langle\hat{R}(\alpha,\beta,\gamma)\rangle =
 \langle\hat{R}(\gamma,\beta,\alpha)\rangle^*\\ 
\label{sym2}
&=& \!\! \langle\hat{R}(\pi-\alpha,\pi-\beta,\gamma)\rangle^* \!\!=\!\!
 \langle\hat{R}(\alpha,\pi-\beta,\pi-\gamma)\rangle^*\\ 
\label{sym3}
&=&
 \langle\hat{R}(\alpha+2\pi,\beta,\alpha)\rangle
=  \langle\hat{R}(\alpha,\beta,\gamma+2\pi)\rangle.
\end{eqnarray}
and are attributed to the properties possessed by the principal-axis cranked
HFB states \cite{HHR82} such as the signature.
\begin{figure*}
\begin{tabular}{ccc}
  \includegraphics[width=0.33\textwidth]{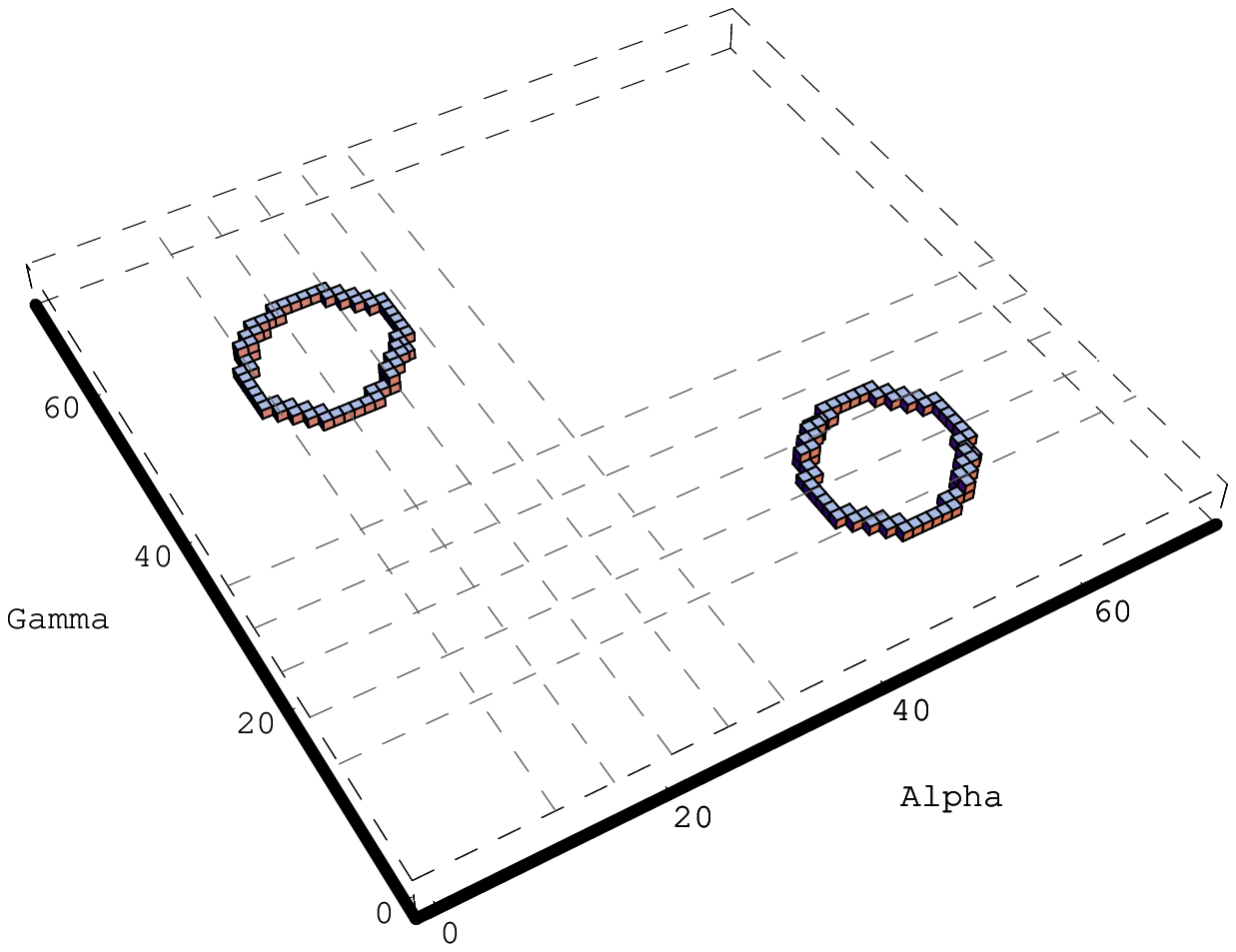}&
  \includegraphics[width=0.33\textwidth]{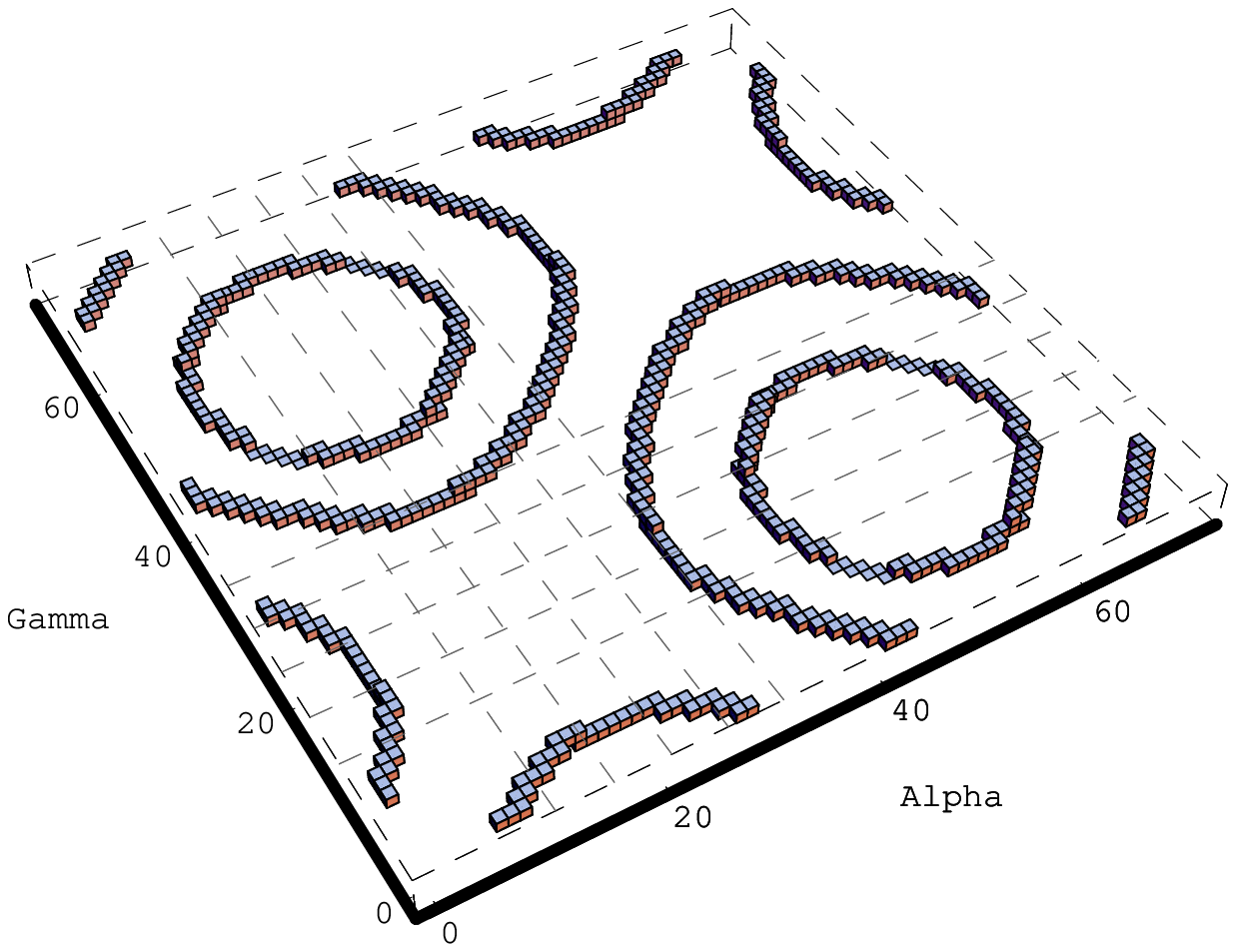}&
  \includegraphics[width=0.33\textwidth]{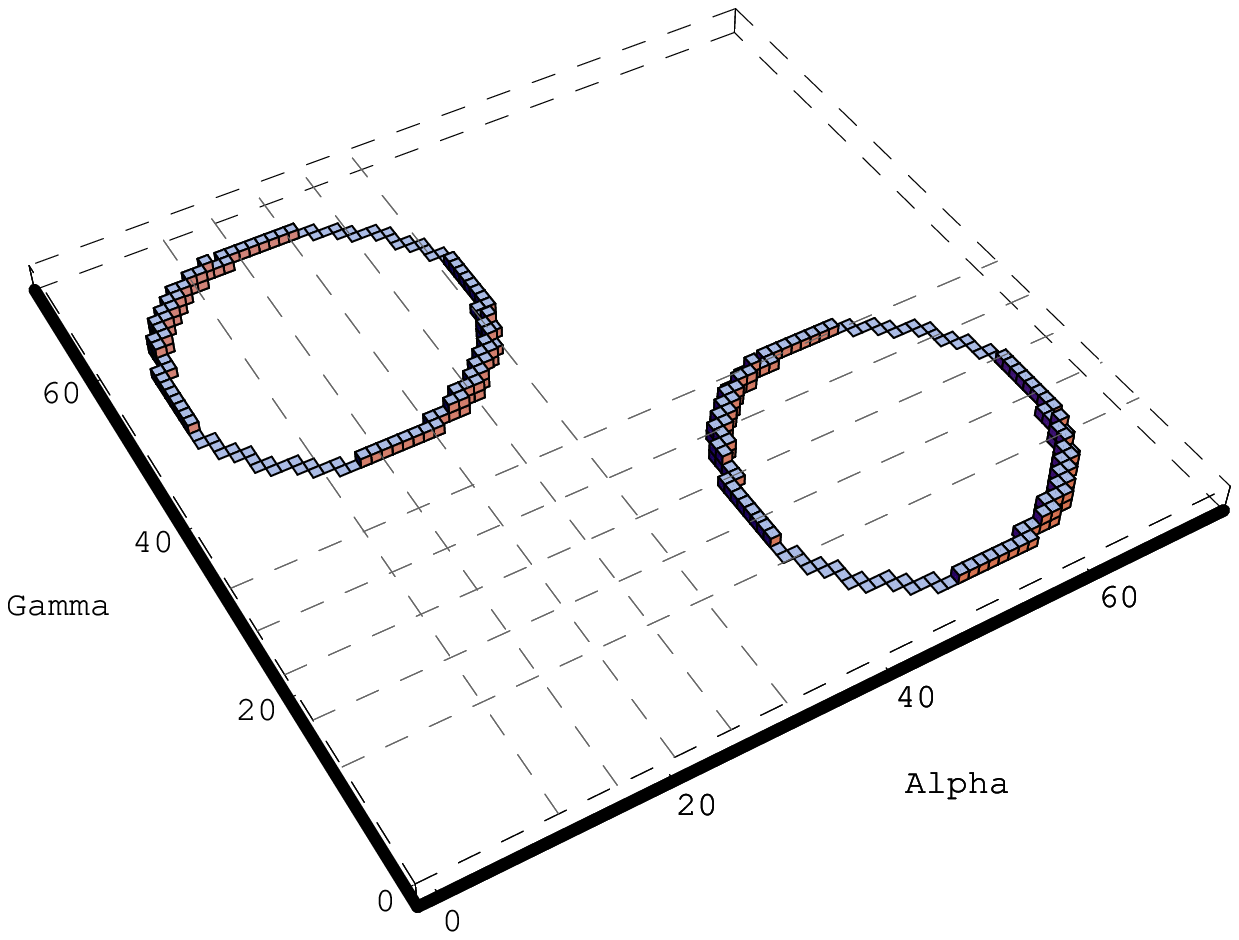}\\
  (a) \quad $\beta\simeq 100^{\circ}$ & 
  (b) \quad $\beta\simeq 110^{\circ}$ & 
  (c) \quad $\beta\simeq 120^{\circ}$\\
\end{tabular}
\caption{Detailed pictures of the essential nodal rings 
in Fig.\ref{fig_nodal2}(B) with respect to $\beta$.}
\label{fig_nodal4}
\end{figure*}
As a result of these symmetries (especially, 
Eqs. (\ref{sym1}) and (\ref{sym2})), there are four identical sets
of nodal lines in Figs.\ref{fig_nodal1} and \ref{fig_nodal2}.
In other words,
there is essentially only one nodal line (ring) out of
the four shown in Figs.\ref{fig_nodal1} and \ref{fig_nodal2}(A),
 while there are sixteen nodal rings
in Figs.\ref{fig_nodal2}(B)
but just four of them are essential.
Two out of these four essential rings form a concentric circles,
as shown in the second portion of Fig.\ref{fig_nodal4}(b). 
(Note the symmetry on the boundary is given in Eqs.(\ref{sym3}).)
Although these rings look to be completely two-dimensional, that is,
confined in the constant $\beta$ plane, one can tell their three-dimensional
structure with a careful look at the figures.

Let us now explain the details of the new approach to determine
the sign $\sigma(\Omega)$ below.
First of all, the third order derivatives of the norm overlap kernels
are chosen to be a measure of the ``smoothness'' of the kernels.
In fact, such a quantity is practically easily computed in our calculations
where the norm overlap kernel and its first order derivatives are 
already obtained.
Using the Taylor expansions for $N(\Omega+\Delta\Omega)$ and
$\frac{\partial N(\Omega+\Delta\Omega)}{\partial\Omega}$,
the third derivative between two points $\Omega$ and 
$\Omega'=\Omega+\Delta\Omega$
is approximately obtained up to an order of ${\cal O}(\Delta \Omega)$,
which is denoted here as $D_3(\Omega',\Omega)$,
\begin{eqnarray}
D_3(\Omega',\Omega) &=&
\frac{6}{(\Delta \alpha)^2}
  \left\{ 
    \frac{\partial N}{\partial \alpha}(\Omega')+
    \frac{\partial N}{\partial \alpha}(\Omega)
  \right\}\\ \nonumber &-&
\frac{12}{(\Delta \alpha)^3}
  \left\{  N(\Omega')-N(\Omega) \right\}.
\end{eqnarray}
The above expression corresponds to a case when 
two points are displaced along the $\alpha$ coordinate, 
i.e., $\Omega=(\alpha,\beta,\gamma)$ and $\Omega'=(\alpha+\Delta
\alpha,\beta,\gamma)$.
Similar expressions hold for a pair of adjacent points displaced 
in the $\beta$ or $\gamma$ direction.
Note that $D_3$ is a complex quantity in contrast to
the residue $\delta\theta$, which is a real quantity.
An advantage to extend the residue $\delta\theta$ 
to the complex $D_3$ is that the continuity is checked 
not only by $\theta(\Omega)$ but also by $r(\Omega)$.
$D_3$ is particularly useful in the vicinity of the nodal lines
where $\ln r(\Omega)$ diverges so quickly. 
In this case, $\delta\theta$ can also change rapidly 
to return the large value as if the boundary
of domains appeared to be detected.
With all the available information
on the continuity both in the norm and argument, 
the boundary between domains $D_n$ and $D_{n+1}$ is
determined with more certainty even in the vicinity of the nodal lines.
Next, starting 
from $\Omega=(0,0,0)$, where $\sigma=1$ is assumed, we determine one by
one the sign of $\sigma$ for the adjacent mesh points so that $|D_3|$
takes a smaller value.
As we proceed, 
conflicting assignments in sign are seen for a point $\Omega=(\alpha,\beta,\gamma)$.
In other words, when $D_3$ is
evaluated along two different links, say 
links between $\Omega$ and 
$\Omega+\Delta\Omega=(\alpha\pm\Delta\alpha,\beta,\gamma)$,
the signs at $\Omega$ are sometimes inconsistent.
In this case, we use ``reliability'' to decide which sign assignment
has priority. Although there is no unique way to define the 
``reliability'', our choice for the definition is
 the ratio of $|D_3|$ between the sign-flipped and unflipped cases 
(the smaller value should be in the denominator): 
larger values of this ratio indicate more reliability.  
This procedure is performed for all the links 
and optimisation on $D_3$ is carried out.
The whole optimisation procedure is 
repeated many times until the average reliability is converged.
The essential point in this new method is that
we attempt to proceed with the analytical continuation through the only links
where a {\it reliable} determination
of the relative sign seems possible.
With this requirement, many problematic points like those near the nodal lines
are circumvented or isolated from the analytical continuation procedure.
\begin{figure}[htbp]
\includegraphics[width=0.5\textwidth,height=0.48\textwidth]{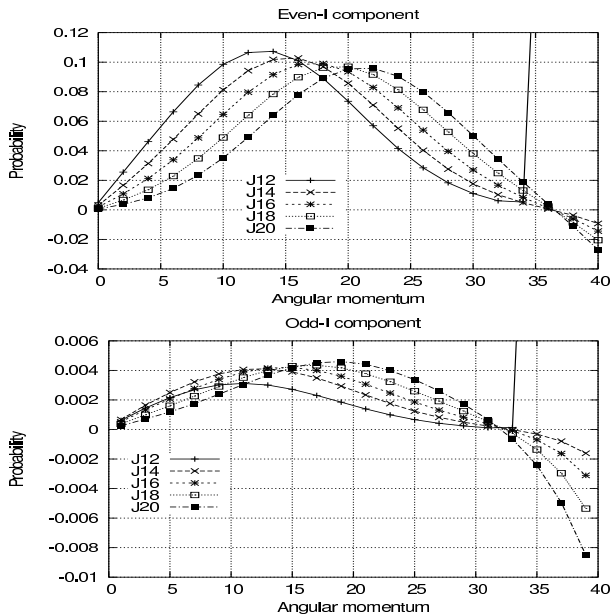}
\caption{Probability distributions for $J=12\hbar$ - $20\hbar$
 with the modified phase determination  method. The upper (lower) panel
shows the profile for even-$I$ (odd-$I$) components.}
\label{prob_mod}
\end{figure}

This new method is quite powerful especially 
when one attempts to determine the sign 
of norm overlap kernels in the presence of the nodal lines.
An advantage of the method is 
that we do not need to find the nodal lines explicitly.
To confirm the advantage,
 we analysed where the disagreements in the sign determination happen 
between the previous naive method (in which the signs are determined one by one
in a scheduled simple order along straight lines)
and the present new method. We then found that the discrepancies are 
concentrated in the vicinity of regions where nodal lines exist.
The improvement in the sign determination by the new method
is clearly demonstrated in Fig.\ref{fig_prob} as the recovery
of the positive definiteness of the probability in the odd-$I$ distributions
of $W^I$.
A disadvantage of this new method is that it is more time-consuming
than the previous method. However, we found that the computation time has been
greatly improved with the combination of the two methods: the initial
sign determination is made by the first naive approach and subsequently
the new optimisation method is carried out.

In the case of $^{170}$Dy, we have checked that this newly developed method
improves the sign determination up to $J=14\hbar$.
Although several doubtful sign assignments start to occur at $J=16\hbar$
even for the higher resolution, 
these potential mistakes do not seem to affect positive
definiteness in probabilities ($|C^I_M|^2\ge 0$) and the sum rule, 
\begin{equation}
\sum_{I=0}W^I = 1,
\end{equation}
up to $J=20\hbar$.
In Fig.\ref{prob_mod}, the results of the 
modified probability distributions are plotted. 
The integrals of each probability distribution curve, 
i.e., the probability sum rule, are presented in Table
\ref{tbl}. 
\begin{table*}[thbp]
\begin{tabular*}{0.5\textwidth}{@{\extracolsep{\fill}}p{2.5cm}ccccc}
\hline
$J \equiv \langle\hat{J}_1\rangle_{\rm HFB}$ & 
$12\hbar$ &
$14\hbar$ &
$16\hbar$ &
$18\hbar$ &
$20\hbar$ \\
\hline
$P$ &
0.998 &
0.991 &
0.981 &
0.965 &
0.939 \\
\hline
\end{tabular*}
\caption{The total probability sum $\displaystyle P=\sum_{I<35\hbar}W^I$
shown in  Fig.\ref{prob_mod}.}
\label{tbl}
\end{table*}
$W^I $ satisfies
the sum rule quite well up to $I=34\hbar$. 
The problem of negative probability does not appear
until $I=31\hbar$. Beyond $I=31\hbar$ for the odd-$I$ components
(and beyond $I=34\hbar$ for the even-$I$), all $W^I$ become  negative.
The reason for this behaviour is attributed to the discretisation
approximation in the integration. With $\Delta\alpha=\Delta\gamma=5^{\circ}$ 
and $\Delta\beta=1^{\circ}$, the integration
in angular momentum projection manages to be valid up to 
\begin{equation}
I\simeq \frac{1}{3}\left(
\frac{2\hbar}{\Delta\alpha}+\frac{2\hbar}{\Delta\beta}+\frac{2\hbar}{\Delta\gamma}\right)= 32\hbar.
\end{equation}
The divergence in the curves of $J=12\hbar$ (beyond $I > 34\hbar$ for
the even-$I$ graph and $I>33\hbar$ for odd-$I$)
comes from the same reason.
To overcome even these difficulties, a finer mesh size should be employed.
Such a line of investigation is now in progress.

In conclusion, we have studied the distribution of zeros of norm
overlap kernels of cranked HFB states. 
Such zeros form one dimensional structure in
the three-dimensional space defined by the Euler angles, so that we
call it a ``nodal line''. 
The existence of nodal lines is numerically demonstrated for the first
time, and we have found they emerge
as spin increases and their structure becomes more complicated as
angular momentum becomes larger ($J > 10\hbar$).
We have shown that it is important to know the
distributions  in order to correct wrong assignments for the sign of norm
overlap kernels. 
The correction is essential for angular momentum projection  to
be performed  with high accuracy for cranked mean-field states.

\begin{acknowledgements}
We greatly thank Professor N. Onishi for giving us good insights
through discussions in order to tackle the present problem.
M.O. is grateful for discussions with Professors H. Flocard
and P.-H. Heenen. Careful reading of the manuscript by Prof. P. Walker
is acknowledged.
Financial support from the Japanese Society for
the Promotion of Sciences (JSPS) 
and an EPSRC advanced research fellowship GR/R75557/01
are appreciated by M.O. 
Parts of the numerical calculations 
were performed at the Centre for Nuclear Sciences (CNS), University of Tokyo,
which is also acknowledged.
\end{acknowledgements}

\end{document}